\documentclass{llncs}
\usepackage{amsmath}
\usepackage{amssymb}
\usepackage{enumerate}
\usepackage{varioref}
\usepackage{charter,eulervm}

\title{{\sc A Note on Triangle Partitions}}

\author{
 Ton~Kloks\inst{} 
}
\institute{
 Department of Computer Science\\
 National Tsing Hua University, Taiwan\\
 {\tt kloks@cs.nthu.edu.tw}
}

\begin{document}

\maketitle

\begin{abstract}
Koivisto studied the partitioning of sets of bounded cardinality. 
We improve his time analysis somewhat, for the special case of 
triangle partitions, and obtain a slight improvement.  
\end{abstract}

\section{Introduction}

Let $G=(V,E)$ be a graph. The triangle partition problem asks 
to partition the vertices of $G$ into vertex-disjoint triangles. 

\bigskip 

The triangle partition problem is NP-complete, even for graphs 
with maximal degree at most four~\cite{kn:rooij}. 
Via the inclusion-exclusion method the problem can 
be solved in $O^{\ast}(2^n)$ time and polynomial space~\cite{kn:bjorklund}. 

\bigskip 

Koivisto analyzes the time complexity of partitioning a 
set into subsets of bounded cardinality. 
His method shows that 
the triangle partition problem can be solved in $O^{\ast}(1.7693^n)$. 

\bigskip 

In this note we simplify his analysis somewhat, for the 
case of triangle partitions, and obtain an $O^{\ast}(1.7549^n)$ 
algorithm. 

\section{Triangle partitions}

Let $G=(V,E)$ be a graph. Let $V=\{1,\dots,n\}$. 
When $G$ has a triangle partition then $q=\frac{n}{3}$ is integer. 

\bigskip 

Koivisto's idea is to search for triangles that are in lexicographic 
order. 

\begin{lemma}
\label{lexicograph triangles}
Consider the lexicograph ordering of the
triangles in a triangle partition $\mathcal{P}$ of $G$.
Then, for $j\in\{1,\dots,q\}$,
the first $j$ triangles in $\mathcal{P}$ must contain
the vertices of $\{1,\dots,j\}$.
\end{lemma}
\begin{proof}
Since the triangles of $\mathcal{P}$ partition $V$
each element of $V$ is in some triangle of $\mathcal{P}$.
Since the triangles are lexicographically ordered, the
first $j$ triangles of $\mathcal{P}$ must contain
all vertices of $\{1,\dots,j\}$.
\qed\end{proof}

\bigskip 

It follows that 
the first $j$ triangles must 
contain $\{1,\dots,j\}$ and the remaining vertices of these 
triangles are $2j$ vertices from $\{j+1,\dots,n\}$. 
Koivisto shows that a dynamic programming 
algorithm can be obtained that runs in time proportional to 
\[\sum_{k=1}^{n/3} \; \binom{n-k}{2k}.\] 

\bigskip 

Via the Hoeffding bound Koivisto obtains his (general) result. 
In the following lemma we simplify the analysis and obtain 
a slightly better bound for the case of triangle partitions. 
  
\begin{lemma}
\label{bound R}
\[\sum_{k=1}^{n/3} \binom{n-k}{2k} = O^{\ast}(1.7549^n).\]
\end{lemma}
\begin{proof}
Let $q=\frac{n}{3}$. 

\medskip 

\noindent
The binomial coefficients can be bounded as follows.
Write
\begin{equation}
\label{iteqn5}
k = \alpha n \quad\text{and}\quad  \beta=\frac{2 \alpha}{1-\alpha}.
\end{equation}
First consider the tails. When $\beta \leq \frac{1}{3}$ then
$\alpha \leq \frac{1}{7}$ and we have
\[\binom{n-k}{2k} \leq 2^{2n/7} \leq 1.22^n.\]
When $\beta \geq \frac{2}{3}$ then $\alpha \geq \frac{1}{4}$ and we find
\[\binom{n-k}{2k} = \binom{n-k}{n-3k} \leq \binom{n}{n-3k} \leq 2^{n/4}
\leq 1.2^n.\]

\medskip

\noindent
Consider the case where $\frac{1}{3} < \beta < \frac{2}{3}$,
that is, $\frac{1}{7} < \alpha < \frac{1}{4}$.
Write $\beta=\frac{1+\epsilon}{2}$.
Then $-\frac{1}{3} < \epsilon < \frac{1}{3}$.
With Stirling's formula we obtain (neglecting polynomial factors)
\begin{equation}
\label{itpeqn1}
\binom{n-k}{2k} \sim \exp \left[\frac{2n}{5+\epsilon}
(\; 2 \ln(2) - (1-\epsilon)ln(1-\epsilon) -
(1+\epsilon)\ln(1+\epsilon)\;)
\right].
\end{equation}

\medskip

\noindent
Now write $\gamma=\frac{1-\epsilon}{2}$.
Then $\frac{1}{3} < \gamma < \frac{2}{3}$
and~(\ref{itpeqn1}) becomes
\begin{equation}
\label{itpeqn2}
\binom{n-k}{2k} \sim \exp\left[ \frac{2n}{3-\gamma} f(\gamma)\right]
\quad\text{where}\quad
\gamma=\frac{1-3\alpha}{1-\alpha} \quad\text{and}\quad \alpha=\frac{k}{n}
\end{equation}
and where $f(\gamma)=-\gamma \ln(\gamma) - (1-\gamma) \ln(1-\gamma)$ is the
entropy function.\index{entropy function}

\medskip

\noindent
Define
\begin{equation}
\label{iteqn3}
g(\gamma)=\frac{1}{3-\gamma} \cdot f(\gamma).
\end{equation}
The function $g(\gamma)$ has its maximum at
$\gamma \leq 0.56985$ and
with this value~(\ref{itpeqn2}) becomes
\begin{equation}
\label{itpeqn4}
\binom{n-k}{2k} =O^{\ast}(1.7549^n).
\end{equation}
\qed\end{proof}

\begin{theorem}
There exists an $O^{\ast}(1.7549^n)$ time 
algorithm that check if the vertices of a graph 
can be partitioned into vertex-disjoint triangles. 
\end{theorem}
\begin{proof}
This follows from Lemma~\ref{bound R} and 
Koivisto's dynamic programming algorithm. 
\qed\end{proof}


\begin{thebibliography}{99}

\bibitem{kn:bjorklund}Bj\"orklund,~A., T.~Husfeldt and M.~Koivisto, 
Set partitioning via inclusion-exclusion, 
{\em SIAM Journal on Computing\/} {\bf 39} (2009), pp.~546--563. 

\bibitem{kn:koivisto}Koivisto,~M.,
Partitioning into sets of bounded cardinality, 
{\em Proceedings of the $4^{\mathrm{th}}$ Workshop on 
Parameterized and Exact Computation\/}, 
Springer-Verlag, LNCS~5917 (2009), pp.~258--263. 

\bibitem{kn:rooij}van~Rooij,~J., M.~van~Kooten~Niekerk and 
H.~Bodlaender, 
Partitioning sparse graphs into triangles -- 
relations to exact satisfiability and very fast exponential time 
algorithm. Technical report Utrecht University, UU-CS-2010-005, 2010. 

\end{thebibliography}
\end{document}